\documentclass[aps,prl,superscriptaddress,showpacs,floatfix,twocolumn]{revtex4}

\usepackage{graphicx}   

\begin{document}

\title{Azimuthal Angle Correlations for Rapidity
Separated Hadron Pairs in $d+Au$ Collisions at $\sqrt{s_{NN}}$ = 200\,GeV}

\newcommand{\abilene}{Abilene Christian University, Abilene, TX 79699, U.S.}
\newcommand{\acadsin}{Institute of Physics, Academia Sinica, Taipei 11529, Taiwan}
\newcommand{\banaras}{Department of Physics, Banaras Hindu University, Varanasi 221005, India}
\newcommand{\barc}{Bhabha Atomic Research Centre, Bombay 400 085, India}
\newcommand{\bnl}{Brookhaven National Laboratory, Upton, NY 11973-5000, U.S.}
\newcommand{\caucr}{University of California - Riverside, Riverside, CA 92521, U.S.}
\newcommand{\ciae}{China Institute of Atomic Energy (CIAE), Beijing, People's Republic of China}
\newcommand{\cns}{Center for Nuclear Study, Graduate School of Science, University of Tokyo, 7-3-1 Hongo, Bunkyo, Tokyo 113-0033, Japan}
\newcommand{\colorado}{University of Colorado, Boulder, CO 80309, U.S.}
\newcommand{\columbia}{Columbia University, New York, NY 10027 and Nevis Laboratories, Irvington, NY 10533, U.S.}
\newcommand{\dapnia}{Dapnia, CEA Saclay, F-91191, Gif-sur-Yvette, France}
\newcommand{\debrecen}{Debrecen University, H-4010 Debrecen, Egyetem t{\'e}r 1, Hungary}
\newcommand{\elte}{ELTE, E{\"o}tv{\"o}s Lor{\'a}nd University, H - 1117 Budapest, P{\'a}zm{\'a}ny P. s. 1/A, Hungary}
\newcommand{\fsu}{Florida State University, Tallahassee, FL 32306, U.S.}
\newcommand{\gsu}{Georgia State University, Atlanta, GA 30303, U.S.}
\newcommand{\hiroshima}{Hiroshima University, Kagamiyama, Higashi-Hiroshima 739-8526, Japan}
\newcommand{\ihepprot}{IHEP Protvino, State Research Center of Russian Federation, Institute for High Energy Physics, Protvino, 142281, Russia}
\newcommand{\illuiuc}{University of Illinois at Urbana-Champaign, Urbana, IL 61801, U.S.}
\newcommand{\isu}{Iowa State University, Ames, IA 50011, U.S.}
\newcommand{\jinrdubna}{Joint Institute for Nuclear Research, 141980 Dubna, Moscow Region, Russia}
\newcommand{\kek}{KEK, High Energy Accelerator Research Organization, Tsukuba, Ibaraki 305-0801, Japan}
\newcommand{\kfki}{KFKI Research Institute for Particle and Nuclear Physics of the Hungarian Academy of Sciences (MTA KFKI RMKI), H-1525 Budapest 114, POBox 49, Budapest, Hungary}
\newcommand{\korea}{Korea University, Seoul, 136-701, Korea}
\newcommand{\kurchatov}{Russian Research Center ``Kurchatov Institute", Moscow, Russia}
\newcommand{\kyoto}{Kyoto University, Kyoto 606-8502, Japan}
\newcommand{\labllr}{Laboratoire Leprince-Ringuet, Ecole Polytechnique, CNRS-IN2P3, Route de Saclay, F-91128, Palaiseau, France}
\newcommand{\lawllnl}{Lawrence Livermore National Laboratory, Livermore, CA 94550, U.S.}
\newcommand{\losalamos}{Los Alamos National Laboratory, Los Alamos, NM 87545, U.S.}
\newcommand{\lpc}{LPC, Universit{\'e} Blaise Pascal, CNRS-IN2P3, Clermont-Fd, 63177 Aubiere Cedex, France}
\newcommand{\lund}{Department of Physics, Lund University, Box 118, SE-221 00 Lund, Sweden}
\newcommand{\muenster}{Institut f\"ur Kernphysik, University of Muenster, D-48149 Muenster, Germany}
\newcommand{\myongji}{Myongji University, Yongin, Kyonggido 449-728, Korea}
\newcommand{\nagasaki}{Nagasaki Institute of Applied Science, Nagasaki-shi, Nagasaki 851-0193, Japan}
\newcommand{\newmex}{University of New Mexico, Albuquerque, NM 87131, U.S. }
\newcommand{\nmsu}{New Mexico State University, Las Cruces, NM 88003, U.S.}
\newcommand{\ornl}{Oak Ridge National Laboratory, Oak Ridge, TN 37831, U.S.}
\newcommand{\orsay}{IPN-Orsay, Universite Paris Sud, CNRS-IN2P3, BP1, F-91406, Orsay, France}
\newcommand{\peking}{Peking University, Beijing, People's Republic of China}
\newcommand{\pnpi}{PNPI, Petersburg Nuclear Physics Institute, Gatchina, Leningrad region, 188300, Russia}
\newcommand{\riken}{RIKEN (The Institute of Physical and Chemical Research), Wako, Saitama 351-0198, JAPAN}
\newcommand{\rikjrbrc}{RIKEN BNL Research Center, Brookhaven National Laboratory, Upton, NY 11973-5000, U.S.}
\newcommand{\saopaulo}{Universidade de S{\~a}o Paulo, Instituto de F\'{\i}sica, Caixa Postal 66318, S{\~a}o Paulo CEP05315-970, Brazil}
\newcommand{\seoulnat}{System Electronics Laboratory, Seoul National University, Seoul, South Korea}
\newcommand{\stonybrkc}{Chemistry Department, Stony Brook University, SUNY, Stony Brook, NY 11794-3400, U.S.}
\newcommand{\stonycrkp}{Department of Physics and Astronomy, Stony Brook University, SUNY, Stony Brook, NY 11794, U.S.}
\newcommand{\subatech}{SUBATECH (Ecole des Mines de Nantes, CNRS-IN2P3, Universit{\'e} de Nantes) BP 20722 - 44307, Nantes, France}
\newcommand{\tenn}{University of Tennessee, Knoxville, TN 37996, U.S.}
\newcommand{\titech}{Department of Physics, Tokyo Institute of Technology, Oh-okayama, Meguro, Tokyo 152-8551, Japan}
\newcommand{\tsukuba}{Institute of Physics, University of Tsukuba, Tsukuba, Ibaraki 305, Japan}
\newcommand{\vandy}{Vanderbilt University, Nashville, TN 37235, U.S.}
\newcommand{\waseda}{Waseda University, Advanced Research Institute for Science and Engineering, 17 Kikui-cho, Shinjuku-ku, Tokyo 162-0044, Japan}
\newcommand{\weizmann}{Weizmann Institute, Rehovot 76100, Israel}
\newcommand{\yonsei}{Yonsei University, IPAP, Seoul 120-749, Korea}
\newcommand{\deceased}{\dagger}
\affiliation{\abilene}
\affiliation{\acadsin}
\affiliation{\banaras}
\affiliation{\barc}
\affiliation{\bnl}
\affiliation{\caucr}
\affiliation{\ciae}
\affiliation{\cns}
\affiliation{\colorado}
\affiliation{\columbia}
\affiliation{\dapnia}
\affiliation{\debrecen}
\affiliation{\elte}
\affiliation{\fsu}
\affiliation{\gsu}
\affiliation{\hiroshima}
\affiliation{\ihepprot}
\affiliation{\illuiuc}
\affiliation{\isu}
\affiliation{\jinrdubna}
\affiliation{\kek}
\affiliation{\kfki}
\affiliation{\korea}
\affiliation{\kurchatov}
\affiliation{\kyoto}
\affiliation{\labllr}
\affiliation{\lawllnl}
\affiliation{\losalamos}
\affiliation{\lpc}
\affiliation{\lund}
\affiliation{\muenster}
\affiliation{\myongji}
\affiliation{\nagasaki}
\affiliation{\newmex}
\affiliation{\nmsu}
\affiliation{\ornl}
\affiliation{\orsay}
\affiliation{\peking}
\affiliation{\pnpi}
\affiliation{\riken}
\affiliation{\rikjrbrc}
\affiliation{\saopaulo}
\affiliation{\seoulnat}
\affiliation{\stonybrkc}
\affiliation{\stonycrkp}
\affiliation{\subatech}
\affiliation{\tenn}
\affiliation{\titech}
\affiliation{\tsukuba}
\affiliation{\vandy}
\affiliation{\waseda}
\affiliation{\weizmann}
\affiliation{\yonsei}
\author{S.S.~Adler}	\affiliation{\bnl}
\author{S.~Afanasiev}	\affiliation{\jinrdubna}
\author{C.~Aidala}	\affiliation{\columbia}
\author{N.N.~Ajitanand}	\affiliation{\stonybrkc}
\author{Y.~Akiba}	\affiliation{\kek} \affiliation{\riken}
\author{A.~Al-Jamel}	\affiliation{\nmsu}
\author{J.~Alexander}	\affiliation{\stonybrkc}
\author{K.~Aoki}	\affiliation{\kyoto}
\author{L.~Aphecetche}	\affiliation{\subatech}
\author{R.~Armendariz}	\affiliation{\nmsu}
% ...
\collaboration{PHENIX Collaboration} \noaffiliation

\date{\today}

\begin{abstract}
We report on two-particle azimuthal angle correlations between
charged hadrons at forward/backward (deuteron/gold going direction)
rapidity and charged hadrons at mid-rapidity in deuteron-gold
($d+Au$) and proton-proton ($p+p$) collisions at $\sqrt{s_{NN}} =
200$\,GeV. Jet structures are observed in the correlations which we
quantify in terms of the conditional yield and angular width of
away-side partners. The kinematic region studied here samples
partons in the gold nucleus carrying nucleon momentum fraction
$x\sim 0.1$ to $\sim0.01$. Within this range, we find no $x$
dependence of the jet structure in $d+Au$ collisions.
\end{abstract}

\pacs{25.75.Dw, 25.75.Gz}


\maketitle


Observations in $d+Au$ collisions at $\sqrt{s_{NN}} = 200$\,GeV at
the Relativistic Heavy Ion Collider (RHIC) reveal a significant
suppression of hadron production at forward rapidity (deuteron-going
direction) relative to $p+p$ reactions scaled up by the equivalent
number of nucleon-nucleon collisions
($N_{coll}$)~\cite{dAuBRAHMS,dAuSTAR,dAuPHENIX}. This suppression is
observed for hadrons with momentum transverse to the beam direction
over the range $p_T \approx 1.5-4$ GeV/$c$.
In contrast, measurements at
mid-rapidity~\cite{ppg28,stcronin,phoboscronin,brahmscronin} and
backward rapidity~\cite{dAuBRAHMS,dAuSTAR,dAuPHENIX} show a modest
enhancement of the particle yield relative to $N_{coll}$ scaling
over the same $p_T$ range. Particle production at forward rapidity
is sensitive to partons in the gold nucleus which carry a small
nucleon momentum fraction (small Bjorken $x$).The suppression has
generated significant theoretical interest including different
calculational frameworks for understanding the
data~\cite{CGC_many,Vogt_LT,Ivan_dAu_pc,Huang_RC}.

One such framework, the Color Glass Condensate (CGC), attempts to
describe the data in terms of gluon saturation~\cite{CGC_many}. At
small $x$ the probability of emitting an extra gluon is large and
the number of gluons grows in a limited transverse area. When the
transverse density becomes large, partons start to overlap and
gluon-gluon fusion processes start to dominate the parton evolution
in the hadronic wave functions. As a result the gluon density
becomes saturated. Since the nonlinear growth of the gluon density
depends on the transverse size of the system, the effects of gluon
saturation are expected to set in earlier (at higher Bjorken $x$)
for heavy nuclei accelerated at ultra-relativistic energies than for
free nucleons.

In the leading order pQCD framework, a quark or gluon jet with large
transverse momentum produced in a hard scattering process (high
momentum transfer or large $Q^2$) must be momentum balanced by
another quark or gluon jet in the opposite direction but with almost
the same transverse momentum. Thus the azimuthal angle correlation
between particles from the pair of jets (referred to as di-jets) is
characterized by two peak structures separated by 180 degrees.  In
CGC calculations, the momentum to balance a jet may come from a
large multiplicity of gluons in the saturation regime, and thus no
single partner jet may appear on the opposite side~\cite{CGC_jet}.
This effect is analogous to the nuclear M\"{o}ssbauer effect, and is
often referred to as the appearance of mono-jets.  Alternative
calculations, describing the suppression of single hadrons at
forward rapidity in $d+Au$ reactions in terms of leading twist pQCD
effects, predict no such mono-jet
feature~\cite{ivan_private_communication}.

We want to probe this high gluon density regime in $d+Au$ collisions
with relatively high transverse momentum particles at forward
rapidity. Such particles are likely to result from hard-scattering
collisions involving small $x$ partons in the gold nucleus. At small
$x$ the gluon density increases rapidly with $Q^2$ and saturation
effects may be relevant for $x \approx 0.01$ at modest $p_T$.  CGC
calculations~\cite{CGC_jet} predict significant suppression of the
conditional yield and widening of away-side jet azimuthal
correlations between rapidity-separated hadron pairs when one of
those hadrons is at forward rapidity.


In this Letter we report on measurements of two-particle azimuthal
angle correlations between unidentified charged hadrons in $p+p$ and
$d+Au$ collisions at $\sqrt{s_{NN}} = 200$\,GeV.   In our analysis,
the two particles are referred to as the {\it trigger} and {\it
associated} particles. The trigger particle is at forward (
$1.4<\eta<2.0$) or backward ($-2.0<\eta< -1.4$) rapidity and the
associated particle is at mid-rapidity, $|\eta|<0.35$. The particles
are separated by an average pseudorapidity gap $<\!\Delta\eta\!>
\sim 1.5$. The criteria for trigger particles, associated particles
and event selection are described elsewhere~\cite{dAuPHENIX,ppg39}.
The two-particle azimuthal angle correlation technique has been used
extensively by RHIC experiments and is described in detail
elsewhere~\cite{starb2b,ppg33,starlpt,ppg39,ppg32}. In this
technique the azimuthal correlation function is formed from the
angular difference, $\Delta \phi = \phi^{assoc} - \phi^{trig}$,
between each trigger and associated particle pair. Two jet peaks are
normally observed in such correlation functions: the near-side peak
($\Delta \phi \sim 0$) in which the two particles come from the same
jet, and the away-side peak ($\Delta \phi \sim \pi$) in which they
come from the back-to-back jets. In addition to these peaks the
correlation functions usually also have a $\Delta \phi$ independent combinatoric background
contribution which is due to trigger-associated pairs 
from different jets or from non-jet processes.

We can construct separate correlation functions that are sensitive
to partons in the gold nucleus with different Bjorken $x$ ranges. By
choosing trigger particles with $1.0<p_T<5.0$ GeV/$c$ at forward
(backward) rapidity, we sample partons in gold nuclei with $x \sim
0.01(0.1)$.
At $x \sim 0.1$, we do not expect to be sensitive to any saturation
effects, but we may be
sensitive at $x \sim 0.01$~\cite{nuclei_dis}. The comparison in
$d+Au$ reactions between these two cases, as well as with the $p+p$
case, may give insights into possible saturation effects on jet
production and other mechanisms for forward rapidity single particle
suppression. It should be noted that the prediction of mono-jets
in~\cite{CGC_jet} assumes one particle at pseudorapidity $\eta =
3.8$ and one at mid-rapidity, thus demonstrating sensitivity at even
lower $x_{Au}$ ($\sim 10^{-4}$) which is outside the range presented
in this analysis.

Data for this analysis were collected by the PHENIX
experiment~\cite{phenixnim} in the 2003 RHIC $p+p$ and $d+Au$ running period. 
In the case of $d+Au$ collisions, we divide the data
into two centrality (impact parameter) classes based on the number
of hits in the backward-rapidity PHENIX Beam-Beam Counter (BBC,
$-3.9 < \eta < -3.0$). Central (peripheral) collisions comprise
$0-40\%$ ($40-88\%$) of the minimum bias cross section respectively.
Using a Glauber model~\cite{dAuPHENIX} and a simulation of the BBC,
we determine \mbox{$<N_{coll}>$}$= 4.7 \pm 0.4$ for peripheral
collisions and \mbox{$<N_{coll}>$}$= 13.0 \pm 0.9$ for central
collisions.

The trigger particles are measured in the PHENIX muon
spectrometers~\cite{phenixnim}.
In this analysis we only select trigger particles
from $1.4<\mid \eta \mid<2.0$ to obtain homogenous acceptance
in transverse momentum from 1$<p_T<5$~GeV/$c$ and to reduce beam
correlated backgrounds. We identify hadrons, as opposed to muons, in
the muon spectrometers by comparing their momentum and penetration
depth. This hadron identification method is described
elsewhere~\cite{dAuPHENIX}.
It is notable that our trigger hadrons have a  modified composition
(pion/kaon/proton ratio) relative to that at the collision vertex
due to species-dependent nuclear interaction cross sections.
Detailed simulations show that kaons make up $65-90\%$ of positively
charged trigger particles and pions make up $70-90\%$ of negatively
charged trigger particles. The baryon contribution to our trigger
particle sample is negligible. We find the two-particle azimuthal
angle correlations for positively and negatively charged trigger
particles to be consistent and therefore combined the results.
The associated particles are unidentified charged hadrons measured
in the PHENIX central spectrometers~\cite{phenixnim} which cover
$\mid\eta\mid<0.35$ and in this analysis have $0.5 < p_T <
2.5$~GeV/$c$. Standard track selection criteria~\cite{ppg39} are
applied.

For comparison we have also included measurements where trigger
particles and associated particles are both measured in the PHENIX
central spectrometers at mid-rapidity.  The $d+Au$ points for this
comparison are from~\cite{ppg39} and the $p+p$ point is an extension
in $p_T$ of the analysis that was published in~\cite{ppg33}.

We define the azimuthal angle correlation function as:
\begin{equation}
CF = \frac{dN(\Delta\phi)/d(\Delta\phi)}{acc(\Delta\phi)}
\end{equation}
where $dN(\Delta\phi)/d(\Delta\phi)$ is the measured two-particle
distribution and $acc(\Delta\phi)$ is the two-particle acceptance
obtained by an event mixing technique in which we mix trigger
particles with associated particles from different events within the
same centrality and collision vertex category. This correction is
necessary because the PHENIX central arm detector is not azimuthally
symmetric and the pair acceptance varies as a function of
$\Delta\phi$.

\begin{figure}[tb]
\includegraphics[width=1.0\linewidth]{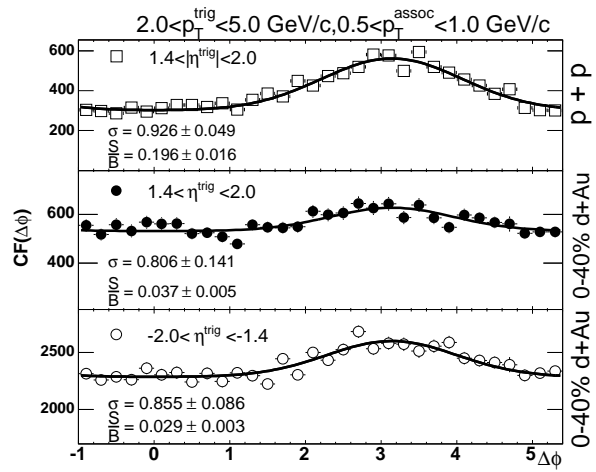}
\caption{\label{fig_pp_dAu_CF} Azimuthal angle correlation
functions.
On the plots, the Gaussian widths from the fits and the signal to
background ratio integrated over $\pi-1 < \Delta \phi < \pi+1$ are
shown. Note that the $y$-axis is zero-suppressed on the middle and
bottom panels.}
\end{figure}

In Figure~\ref{fig_pp_dAu_CF}, we show the correlation functions for
trigger particles with
$p_{T} = 2-5$~GeV/$c$ and associated particles with
$p_{T} = 0.5-1.0$~GeV/$c$. The top panel is for $p+p$ collisions
where we have combined the results from forward and backward
pseudorapidity since the collision system is symmetric. The middle
and bottom panels of Figure~\ref{fig_pp_dAu_CF} show the correlation
functions for central $d+Au$ collisions. The middle panel is for the
trigger particle at forward rapidity and the bottom panel is for the
trigger particle at backward rapidity. A clear peak is seen near
$\Delta\phi=\pi$ in all cases corresponding to the away-side jet. It
is notable that there is no peak near $\Delta\phi=0$, as expected,
because the rapidity gap between the two particles is larger than
the width of the near side jet.

After constructing the correlation functions in various bins in
$p_T^{assoc}$, $p_T^{trig}$ and $\eta^{trig}$ we used two methods to
determine the unnormalized number of trigger-associated particle
pairs, $N_{pair}$, above a constant background. In the first method,
we define
\begin{equation}
N_{pair} = \sum_{\Delta\phi = \pi-1}^{\pi+1}CF(\Delta\phi) -
\sum_{\Delta\phi = -1}^{+1}CF(\Delta\phi),
\end{equation}
where the first term is the integral of the correlation function in
the area of the correlation peak ($\pi-1<\Delta\phi<\pi+1$) and the
second term is the integral away from the peak ($-1<\Delta\phi<1$).
In the second method we fit the correlation function with a Gaussian
distribution centered at $\Delta\phi = \pi$ plus a constant
background.  The values of $N_{pair}$ obtained by each method are
found to be consistent and the small differences are included in our
systematic errors. The solid lines in Figure~\ref{fig_pp_dAu_CF}
show the resulting fits and the Gaussian width parameters ($\sigma$)
together with the integrated signal to background ratios
($\frac{S}{B}$) over the signal region ($\pi-1 < \Delta \phi <
\pi+1$) are quoted.

The conditional yield (per trigger particle ) is defined to be
\begin{equation}\label{eq_CY}
CY = \frac{N_{pair}/\varepsilon_{assoc}}{N_{trig}},
\end{equation}
where $\varepsilon_{assoc}$ ($\sim 0.15 \pm 0.015$) is the efficiency
times acceptance for associated particles and $N_{trig}$ is the
number of trigger particles used to generate the correlation
function. $\varepsilon_{assoc}$ is obtained for each colliding
system, centrality class, and transverse momentum bin by a GEANT
based simulation of the PHENIX detector~\cite{ppg39}.

\begin{figure}[tb]
\includegraphics[width=1.0\linewidth]{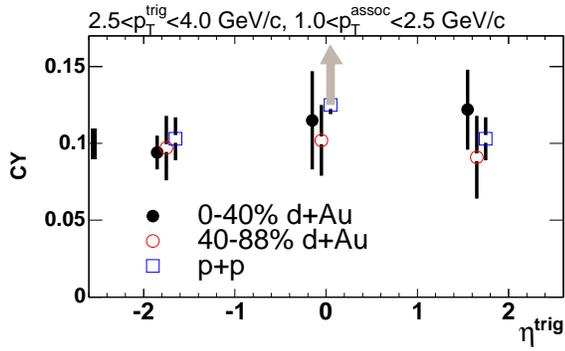}
\caption{\label{fig_yield} Conditional yields are shown as a
function of trigger particle pseudorapidity. The data points at
mid-rapidity for $d+Au$ collisions are from~\cite{ppg39}. To
increase visibility, we artificially shift the data points belonging
to the same $\eta^{trig}$ bin. The errors on each points are
statistical errors. The black bar around $0.1$ on the left of the
plot indicates a $10\%$ common systematic error for all the data
points due to the determination of associated particle efficiency.
There is an additional $+0.037$ systematic error on the mid-rapidity
$p+p$ point from jet yield extraction, which is shown as the arrow
on that point (similar analysis as~\cite{ppg33}).}
\end{figure}

It is interesting to plot the conditional yields as a function of
$\eta^{trig}$. Changing $\eta^{trig}$ from $-2.0$ to $2.0$
effectively changes the range of Bjorken $x$ of sampled partons in
gold nuclei from $0.1^{+0.06}_{-0.04}$ to $0.01^{+0.02}_{-0.007}$.
Results are shown in Figure~\ref{fig_yield}. The first observation
is that there is no difference beyond statistical fluctuations in
the conditional yields for $p+p$, $d+Au$ peripheral, or $d+Au$
central collisions at any trigger particle pseudorapidity.

\begin{figure}[tb]
\includegraphics[width=1.0\linewidth]{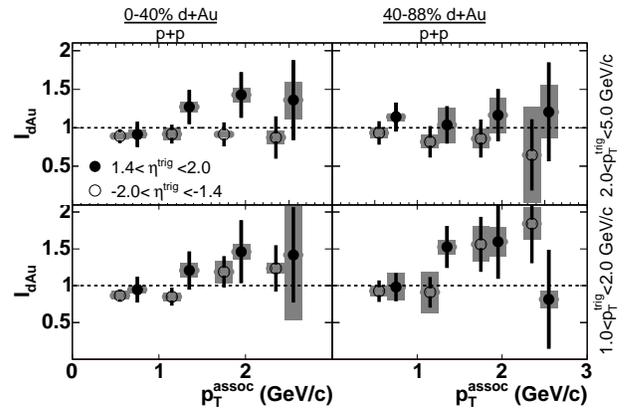}
\caption{\label{fig_corr_RdA} $I_{dAu}$
  vs.~$p_T^{assoc}$ for different centrality, $p^{trig}_T$ and
  $\eta^{trig}$ bins. To increase visibility, we artificially shift
  the data points belonging to the same
$p_T^{assoc}$ bin.}
\end{figure}

We further quantify any nuclear modification in the conditional
yield by defining the following ratio:
\begin{equation}
I_{dAu} = \frac{CY\mid_{d+Au}}{CY\mid_{p+p}}
\end{equation}
Figure~\ref{fig_corr_RdA} shows the ratio $I_{dAu}$ as a function of
$p^{assoc}_T$ for central and peripheral $d+Au$ collisions, two
different $p^{trig}_T$ ranges, and for forward and backward rapidity
trigger particles. In the plot, the shaded bands on each of the data
points are the systematic errors which are the differences in
$N_{pair}$ obtained from the two methods described above and are
independent from point-to-point. There is also a point-by-point
correlated $\sim 2$\% systematic uncertainty in the centrality
dependence of $\varepsilon_{assoc}$ determined by embedding Monte
Carlo tracks into real events. The size of this uncertainty is
comparable to the width of the $I_{dAu}=1$ line.

Our measurement of the nuclear modification of the conditional yield
indicates that $I_{dAu}$ with the trigger particle at forward
rapidity (sampling low-$x$ partons in the gold nucleus) and backward
rapidity (sampling high-$x$ partons in the gold nucleus) are both
consistent with one. There may even be some evidence of slight
enhancement for the case with trigger particles at forward
pseudorapidity in central $d+Au$ collisions.  We note that if
mono-jets were a major contributor to our trigger particle sample in
our $x$ range, we would have expected a decrease in the conditional
yield for $d+Au$ central collisions when the trigger particle is at
forward pseudorapidity. Our measurement is inconsistent with any
large nuclear suppression ({\it i.e.} mono-jets) of the jet
structure in this kinematic range. However, we note that in these
modest $p_T$ ranges, there may be contributions from both ``hard''
(large $Q^2$) processes and ``soft'' coherent (small $Q^2$)
processes.  In $d+Au$ collisions ``soft'' particle production is
shifted away from forward rapidity towards backward
rapidity~\cite{PHOBOS_dAu}.  Thus, the fraction of hadrons at
forward rapidity from ``hard'' processes may be increased in central
$d+Au$ reactions, which may offer an explanation for the modest
enhancement seen in the conditional yield for this case and could
also mask off small mono-jet signals.

\begin{figure}[tb]
\includegraphics[width=1.0\linewidth]{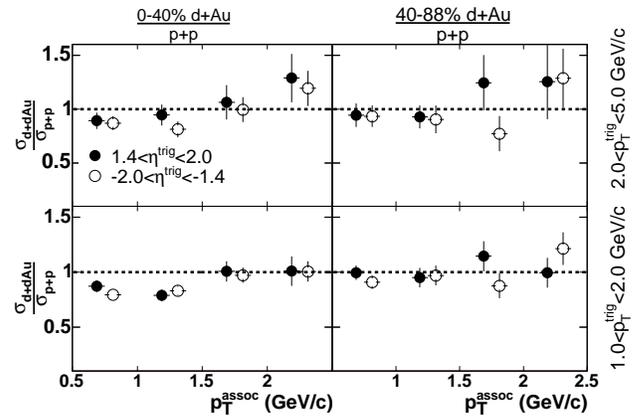}
\caption{\label{fig_width_ratio} The ratio of correlation peak
widths between $d+Au$ and $p+p$ collisions.
Only statistic error is shown. To increase visibility, we
artificially shift the data points belonging to the same
$p_T^{assoc}$ bin. }
\end{figure}

We have also compared the Gaussian widths of the correlation peaks
in $d+Au$ collisions to the widths in $p+p$ collisions. The ratios
of the $d+Au$ widths to the $p+p$ widths are plotted in
Figure~\ref{fig_width_ratio} as a function of $p_T^{assoc}$.   There
may be a hint of a slight $p_T^{assoc}$ dependence in the ratio, but
overall there is no significant difference in the width in $d+Au$
collisions for different $\eta^{trig}$.

In conclusion, we have measured the two-particle azimuthal angle
correlations in two centrality categories of $d+Au$ collisions and
in $p+p$ collisions. We measured trigger particles at forward,
backward and mid-rapidity and correlated them with associated
particles at mid-rapidity. The associated particle conditional
yields in central $d+Au$ collisions are consistent with the
conditional yield in $p+p$ collisions. These conditional yields also
do not change as we vary the trigger particle pseudorapidity over
the range $ \mid \eta \mid <2.0$. We have also compared the widths
of the away-side jet peaks in $d+Au$ and in $p+p$ collisions, and
find no evidence for large $\eta^{trig}$-dependent modification.


We thank the staff of the Collider-Accelerator and Physics
Departments at BNL for their vital contributions.  
We acknowledge support from 
the Department of Energy and NSF (U.S.A.), 
MEXT and JSPS (Japan), 
CNPq and FAPESP (Brazil), 
NSFC (China), 
IN2P3/CNRS, CEA, and ARMINES (France), 
BMBF, DAAD, and AvH (Germany), 
OTKA (Hungary), 
DAE and DST (India), 
ISF (Israel), 
KRF and CHEP (Korea), 
RMIST, RAS, and RMAE (Russia), 
VR and KAW (Sweden), 
U.S. CRDF for the FSU, 
US-Hungarian NSF-OTKA-MTA, 
and US-Israel BSF.


\def\Journal#1#2#3#4{{#1}{\bf #2}, #3 (#4)}
\def\IJMPA{{Int. J. Mod. Phys.}~{\bf A}}
\def\JPG{{J. Phys}~{\bf G}}
\def\NCA{Nuovo Cimento\ }
\def\NIM{Nucl. Instrum. Methods\ }
\def\NIMA{{Nucl. Instrum. Methods\ }~{\bf A}}
\def\NPA{{Nucl. Phys.}~{\bf A}}
\def\NPB{{Nucl. Phys.}~{\bf B}}
\def\PLB{{Phys. Lett.}~{\bf B}}
\def\PLC{Phys. Repts.\ }
\def\PR{Phys. Rev.\ }
\def\PRL{Phys. Rev. Lett.\ }
\def\PRD{Phys. Rev. {\bf D}\ }
\def\PRC{Phys. Rev. {\bf C}\ }
\def\RMP{Rev. Mod. Phys.\ }
\def\SPJ{Sov. Phys. JETP\ }
\def\SJNP{Sov. J. Nucl. Phys.\ }
\def\ZPC{{Z. Phys.}~{\bf C}}


\begin{references}

\bibitem{dAuBRAHMS}
I.~Arsene {\it et al}., \Journal{\PRL}{93}{242303}{2004}

\bibitem{dAuSTAR}
J.~Adams {\it et al}., \Journal{\PRC}{70}{064907}{2004}

\bibitem{dAuPHENIX}
S.\,S.~Adler {\it et al}., \Journal{\PRL}{94}{082302}{2005}

\bibitem{ppg28}
S.\,S.~Adler {\it et al}., \Journal{\PRL}{91}{072303}{2003}

\bibitem{stcronin}
J.~Adams {\it et al.}, \Journal{\PRL}{91}{072304}{2003}

\bibitem{phoboscronin}
B.B.~Back {\it et al.}, \Journal{\PRL}{91}{072302}{2003}

\bibitem{brahmscronin}
I.~Arsene {\it et al.}, \Journal{\PRL}{91}{072305}{2003}

\bibitem{CGC_many} L. McLerran and R. Venugopalan,
\Journal{\PRD}{49}{2233}{1994};
\Journal{\PRD}{49}{3352}{1994}

\bibitem{Vogt_LT}
R.~Vogt, \Journal{\PRC}{70}{064902}{2004}

\bibitem{Ivan_dAu_pc}
J.~Qiu, I.~Vitev, hep-ph/0410218

\bibitem{Huang_RC}
R.\,C.~Hwa, C.\,B.~Yang and R.\,J.~Fries, \Journal{\PRC}{71}{024902}{2005}

\bibitem{CGC_jet}
D.~Kharzeev, E.~Levin and L.~McLerran,
\Journal{\NPA}{748}{627}{2005}

\bibitem{ivan_private_communication}
Private communication with Ivan Vitev.



\bibitem{ppg39}
S.\,S.~Adler {\it et al}., nucl-ex/0510021

\bibitem{starb2b}
C.~Adler {\it et al.}, \Journal{\PRL}{90}{082302}{2005}

\bibitem{ppg33}
S.\,S.~Adler {\it et al}., \Journal{\PRC}{71}{051902(R)}{2005}

\bibitem{starlpt}
J.~Adams {\it et al}., \Journal{\PRL}{95}{152301}{2005}

\bibitem{ppg32}
S.\,S.~Adler {\it et al}., nucl-ex/0507004


\bibitem{nuclei_dis}
M. Arneodo, Phys. Rep. {\bf 240}, 301 (1994).

\bibitem{phenixnim} K.~Adcox
{\it et al}., \Journal{\NIMA}{499}{469}{2003}






\bibitem{PHOBOS_dAu}
B. B. Back {\it et al}., \Journal{\PRC}{72}{031901}{2005}

\end{references}
\end{document}